\documentclass[aps,prd,preprintnumbers,nofootinbib,preprint]{revtex4}
\usepackage{bm}
\usepackage{latexsym}
\usepackage{dcolumn}
\usepackage{amsmath,amsfonts,amssymb}
\usepackage{graphicx,epsfig}
\usepackage{amsthm}
\usepackage{url,hyperref}

\usepackage{slashed}

\newcommand{\be}{\begin{eqnarray}}
\newcommand{\ee}{\end{eqnarray}}
\newcommand{\bea}{\begin{eqnarray}}
\newcommand{\eea}{\end{eqnarray}}

\def\comment#1{}

\usepackage{color}

\definecolor{darkred}{rgb}{.8,0,0}

\definecolor{darkblue}{rgb}{0,0,.7}

\definecolor{darkgreen}{rgb}{0,.7,0}

\begin{document}

%
%
\title{Joule-Thomson expansion of a specific black hole in different dimensions}
%
%
%
%
%
\author{{Sen Guo}$^{1}$}
\author{Yan Han$^{1}$}
\author{Guo-Ping Li$^{1}$} \email[email:~]{gpliphys@yeah.net}

\affiliation{$^1$College of Physics and Space Science, China West Normal University, Nanchong 637002, China\vspace{1ex}}

%
%
%
%
%
\begin{abstract}
%
%
%
%
%
%
\par\noindent
In this paper, we have studied the Joule-Thomson expansion of a specific black hole in $f(R)$ gravity coupled with Yang-Mills field. Firstly, $4$-dimensional equation of state and the Joule-Thomson coefficient were derived from thermodynamic quantities, the critical temperature and the inversion temperature were obtained, and the isenthalpic curves and the inversion curves have plotted in the $T-P$ plane. Then, by considering the higher dimensions black hole, the inversion temperature was obtained, the isenthalpic curves and the inversion curves also were plotted, and the effect of nonlinear term $\eta$ and dimension for the Joule-Thomson expansion is demonstrated visually by depicting different curves. Finally, we compared the ratio of the inversion temperature to the critical temperatures with the van der Waals fluid and other black hole, which the ratio is not equal to $0.5$, and the dimension changes, the ratio also changes.\\
\end{abstract}
%
%
%
%
\maketitle
%
%
%
%
%
%
\section{Introduction}
\label{intro}
%
%
\par\noindent
As we all known, the research on the thermodynamic system of black holes in general relativity is analogous to the general thermodynamic system in classical physics. The research is based on the four hypotheses of thermodynamics, the zero law and three other laws. Hawking first proposed black hole temperature \cite{Ref1}, it means that the black holes can radiate particles near the event horizon through gravitational interactions. This conclusion makes it very important to study the thermodynamic systems of various types of black hole solutions in general relativity. Subsequently, the thermodynamic properties of the black holes have been widely studied \cite{Ref2,Ref3,Ref4}. Because of the AdS/CFT correspondence, people have been interested in AdS black hole, and its thermodynamic properties have always been investigated. Hawking and Page have been pioneered research the thermodynamic properties of the Schwarzschild-AdS black hole \cite{Ref5}, Chamblin has studied the first-order phase transition of the charged RN-AdS black hole \cite{Ref6,Ref7}. Subsequent studies found that the AdS black holes behave like the van der Waals fluids when they contain charge/rotating \cite{Ref8,Ref9}.
\par
Recently, when studied the thermodynamic properties of AdS black holes, it has been found that the cosmological constant can be considered as the thermodynamic pressure, and conjugate variables as the thermodynamic volumes, which can be used to analogize the charged AdS black holes with the van der Waals gas-liquid systems \cite{Ref10}. At the same time, the change of the cosmological constant is included in the first law of black hole thermodynamics \cite{Ref11,Ref12,Ref13,Ref14,Ref15}, which can make the first law of black hole thermodynamics consistent with the Smarr relation. And if the cosmological constant is taken as the thermodynamics pressure in the first law of thermodynamics, the mass of the black holes should be interpreted as the enthalpy. Based on these considerations, the thermodynamic properties of black holes in the extended phase space have been extensively studied by used the extended structure of thermodynamic phase space \cite{Ref16,Ref17,Ref18,Ref19,Ref20,Ref21,Ref22,Ref23,Ref24,Ref25,Ref26,Ref27,Ref28,Ref29,Ref30,Ref31,Ref32,Ref33,Ref34,Ref35,Ref36,Ref37,Ref38}.
\par
More recently, the Joule-Thomson expansion has been creatively extended to the charged AdS black holes by \"{O}kc\"{u} and Aydyner \cite{Ref39,Ref40}. In classical thermodynamics, the Joule-Thomson expansion refers to the process of gas expanding from high pressure to low pressure through porous plugs, which is an isenthalpy process. Subsequently this creative work has been extended to all kinds of the black holes, such as d-dimensional charged AdS black holes \cite{Ref41}, quintessence RN-AdS black hole \cite{Ref42}, Holographic superfluid \cite{Ref43}, charged AdS black hole in $f(R)$ gravity \cite{Ref44}, AdS black hole with a global monopole \cite{Ref45}, charged AdS black hole in Lovelock gravity \cite{Ref46}, Gauss-Bonnet black hole \cite{Ref47}, AdS black holes with momentum relaxation \cite{Ref48}, regular(Bardeen)-AdS black hole \cite{Ref49}, charged AdS black holes in Rainbow gravity \cite{Ref50}, Hayward-AdS black hole \cite{Ref51}.

\par
In addition, one hopes to explain a number of cosmological problems covering dark energy, accelerated expansion and quantum gravity and many related matters, those leads to that the extension/revision of general relativity is very interesting. In particular, it shows that $f(R)$ gravity is a extension of general relativity, and it is very important modification of Einstein's theory of gravity, it is modified by adding the higher power of scalar curvature $R$, Riemann tensor and Rich tensor or their derivatives by Lagrange formula, and it can be used to study the accelerated expansion of the universe \cite{Ref52,Ref53,Ref54,Ref55}. On the other hand, we know that the Yang-Mills field is one of the most interesting non-abelian gauge theory, which is used inside the nuclei with short range, the unified theory of weak interaction and electromagnetic interaction can be established by used Yang-Mills field, and this theory provides a powerful tool for studying strong interaction. By studying the string theory model, the Yang-Mills field equation under the low energy limit was found, and Yasskin found the first black hole solution in the theory of Yang-Mills coupled to Einstein theory \cite{Ref56}, this theory possesses ¡°hairy¡± black hole solutions, whose metric is not a member of the Kerr-Newman type \cite{Ref57}. Unlike the Kerr-Newman black holes, although only a few parameters are needed to study gauges and matter fields, the global charge measurable at infinite distances is not the only condition that determines the geometry of the horizon.

\par
In this paper, there are two reasons why we consider the charged AdS black hole in $f(R)$ gravity coupled with Yang-Mills field, one is the conformal invariance, the other is the non-linear parameters contained in this black hole. Our starting point is the nonlinear Yang-Mills field, which is similar to the self-interacting scalar field defined in quantum field theory, and the Lagrange is an arbitrary function of the Yang-Mills invariant. In the previous research, one has been investigated the thermodynamic properties of the RN-AdS black holes in $f(R)$ gravity \cite{Ref16,Ref44}, but when we consider the coupling of non-Abelian gauge field (Yang-Mills field) to $f(R)$ gravity field, the effect of non-linear parameter on the Joule-Thomson of black holes is unknown. Therefore, it is meaningful to study the Joule-Thomson expansion of this kind of coupled black hole.

\par
The paper is organized as follows: In Sec \ref{sec:1}, we will briefly review thermodynamic problems in $f(R)$ gravity coupled with Yang-Mills field. In Sec \ref{sec:2}, we investigate the Joule-Thomson expansion of a special black hole in $f(R)$ gravity coupled with Yang-Mills field in different dimensions. In Sec\ref{sec:5}, we discuss the conclusion.

\section{The thermodynamics of the d-dimensional specific black hole in $f(R)$ gravity coupled with Yang-Mills field}
\label{sec2}
%
%
%
\par\noindent
In this section, we briefly reviewed the thermodynamic quantities of a specific black hole in $f(R)$ gravity coupled with Yang-Mills field. The action of $f(R)$ gravity minimally coupled with Yang-Mills field is \cite{Ref58}
\begin{equation}
S=\int d^{d}x \sqrt{-g}\Big[\frac{f(R)}{16 \pi}+{\pounds}(F)\Big],
\label{2-1}
\end{equation}
the function of Ricci scalar $R$ constitutes $f(R)$ and $\pounds({F})$ is interpreted as lagrangian of the nonlinear Yang-Mills field with $F=\frac{1}{4}tr\Big(F^{(a)}_{\mu\nu}F^{(a)\mu\nu}\Big)$ where
\begin{equation}
F^{(a)}=\frac{1}{2}F^{a}_{\mu\nu}dx^{\mu}\wedge dx^{\nu}.
\label{2-2}
\end{equation}
Here the internal index $(a)$ for the degrees of freedom of the non-abelian Yang-Mills field, and this nonlinear Yang-Mills field come from linear Yang-Mills field (${\pounds}(F)=-\frac{1}{4 \pi}F^{s}$) for $s=1$ and $f_{R}=\frac{df(R)}{dR}=\eta r$ in which $\eta$ is a integration constant. By solving $f(R)$ gravity for Einstein field equation coupled with Yang-Mills field, one can get the spherically symmetric metric is written as \cite{Ref58}
\begin{equation}
ds^2=-f(r)dt^2+\frac{1}{f(r)}dr^2+r^2 d\Omega^{2}_{d-2},
\label{2-3}
\end{equation}
where $f(R)$ is only a function of $r$, and
\begin{equation}
d\Omega^{2}_{d-2}=d\theta_{1}^{2}+\mathop{\sum}_{i=2}^{d-2}\mathop{\prod}_{j=1}^{i-1}\sin^{2}\theta_{j}d\theta^{2}_{i},
\label{2-4}
\end{equation}
with $0\leq \theta_{d-2}\leq 2\pi$, $0\leq \theta_{i}\leq \pi$, $1\leq i\leq d-3$. One metric function $f(r)$ is
\begin{equation}
f(r)=\frac{d-3}{d-2}-\Lambda r^{2}-\frac{M}{r^{d-2}}-\frac{(d-1){(d-2)}^{\frac{d-1}{2}}{(d-3)}^{\frac{d-1}{4}}}{2^{\frac{d-5}{2}}\eta d} \frac{q^{\frac{d-1}{2}}\ln r}{r^{d-2}}.
\label{2-5}
\end{equation}
In those equations $\Lambda=-\frac{1}{l^{2}}$, the $M$ and the $q$ are the AdS radius, the black hole mass, and charge of the black hole. One can obtain black hole event horizon as largest root of $f(r_{+})=0$. The mass of black hole in Eq.(\ref{2-5}) is given by
\begin{equation}
M=\frac{1}{r_{+}^{d-2}}\Big[\frac{d-3}{d-2}-\Lambda r_{+}^{2}-\frac{2^{\frac{5-d}{2}}(d-3)^{\frac{d-1}{4}}(d-2)^{\frac{d-1}{2}}(d-1)q^{\frac{d-1}{2}}r_{+}^{2-d} \ln r_{+}}{\Lambda \eta}\Big],
\label{2-6}
\end{equation}
and the expression for entropy is given by \cite{Ref58}
\begin{equation}
S=\frac{A_{h}}{4}\eta r_{+},
\label{2-7}
\end{equation}
where$A_{h}=\frac{d-1}{\Gamma (\frac{d+1}{2})}{\pi}^{\frac{d-1}{2}}r_{+}^{d-2}$. The Hawking temperature of the black hole is calculated by
\begin{equation}
T=\Big(\frac{\partial M}{\partial S}\Big)_{q,P}.
\label{2-8}
\end{equation}
\par
Then, the thermodynamic pressure in terms of the cosmological constant in the extended phase space \cite{Ref10,Ref11,Ref12}, i.e.
\begin{equation}
P=-\frac{\Lambda}{8 \pi},
\label{2-9}
\end{equation}
and the thermodynamic volume is $V=\Big(\frac{\partial M}{\partial P}\Big)_{S,q}$,
\begin{equation}
V=\frac{\Omega_{d-2}r_{+}^{d-1}\eta}{n-1}.
\label{2-10}
\end{equation}
Moreover, one may consider that the black hole mass is a function of three thermodynamic variable, i.e. $M \equiv M(S,P,q)$. The differential form of mass is
\begin{equation}
dM=\Big(\frac{\partial M}{\partial S}\Big)_{q,P}dS+\Big(\frac{\partial M}{\partial q}\Big)_{S,P}dq+\Big(\frac{\partial M}{\partial P}\Big)_{S,q}dP,
\label{2-11}
\end{equation}
where with the quantity $\Phi=\Big(\frac{\partial M}{\partial q}\Big)_{S,P}$, the generalized Smarr relation corresponding to the first law of thermodynamics of the black holes is
\begin{equation}
M=\frac{d-2}{d-3}TS-\frac{2}{d-3}VP+\Phi q.
\label{2-12}
\end{equation}
In the following section, we will investigate the Joule-Thomson expansion of this black hole.
%
%
%
\section{Joule-Thomson expansion of a specific black hole in $f(R)$ gravity coupled with Yang-Mills field}
\label{sec3}
%
%
%
\par\noindent
In this section, we will discuss the Joule-Thomson expansion of a specific black hole in $f(R)$ gravity coupled with Yang-Mills field, and will investigate on different dimensions.
\subsection{The Joule-Thomson expansion}
\label{sec3-1}
\par
The Joule-Thomson expansion is one of the classical physical processes describing the temperature change of gas from high pressure to low pressure irreversibly through a porous plug or valve. It is mainly describe the process of gas expansion, the process of temperature decrease is called the cold effect/positive effect, the temperature rise is called the heat effect/positive effect, and the enthalpy remains constant during the expansion process in the extended phase space. Based on these properties, one can obtained the Joule-Thomson coefficient $\mu$,
\begin{equation}
\mu=\Big(\frac{\partial T}{\partial P}\Big)_{H}.
\label{3-1-1}
\end{equation}
\par
The cooling-heating regions can be determined by the sign of Eq.(\ref{3-1-1}). Because the pressure always decreases during expansion, the increase or decrease of temperature affects the sign of $\mu$, which the change of temperature is positive (negative) $\mu$ is negative (positive) and so gas warms (cools). In the extended phase space, we compare the black hole system to a van der Waals fluid system with a fixed number of particles, where consider canonical ensemble with fixed change $q$. The Joule-Thomson coefficient is given by \cite{Ref39}
\begin{equation}
\mu=\Big(\frac{\partial T}{\partial P}\Big)_{H}=\frac{1}{C_{P}}\Big[T\Big(\frac{\partial V}{\partial T}\Big)_{P}-V\Big],
\label{3-1-2}
\end{equation}
and setting $\mu=0$, we can obtain the inversion temperature, i.e.
\begin{equation}
T_{i}=V\Big(\frac{\partial T}{\partial V}\Big)_{P}.
\label{3-1-3}
\end{equation}
%
%
\subsection{The Joule-Thomson expansion of a specific black hole in $f(R)$ gravity coupled with Yang-Mills field in different dimensions}
\label{sec3-2}
%
%
\par\noindent
When $d=4$, the metric function $f(r)$ is given by Eqs. (\ref{2-5}) and (\ref{2-9})
\begin{equation}
f(r)=\frac{1}{2}+8 \pi P r^{2}-\frac{M}{r^{2}}-\frac{3q^{3/2}\ln r}{r^2 \eta}.
\label{3-2-1}
\end{equation}
One can obtain black hole event horizon as largest root of $f(r_{+})=0$, and the mass of black hole in Eq. (\ref{3-2-1})
\begin{equation}
M=\frac{\eta r_{+}^{2}+16 \eta P \pi r_{+}^{4}-6q^{3/2}\ln r_{+}}{2\eta}.
\label{3-2-2}
\end{equation}
The temperature is
\begin{equation}
T=\Big(\frac{\partial M}{\partial S}\Big)_{P,q}=\frac{-3 q^{2/3}+\eta r_{+}^{2}+32 \eta P \pi r_{+}^{4}}{4\eta \pi r_{+}^{3}}.
\label{3-2-3}
\end{equation}
According to the expression of temperature Eq.(\ref{3-2-3}), we can get the equation of state
\begin{equation}
P=\frac{T}{8r_{+}}+\frac{3 q^{3/2}}{32 \eta \pi r_{+}^{4}}-\frac{1}{32 \pi r_{+}^{2}}.
\label{3-2-4}
\end{equation}
The critical points obtained from \cite{Ref59}
\begin{equation}
\frac{\partial P}{\partial r_{+}}=0=\frac{\partial^{2}P}{\partial r_{+}^{2}},
\label{3-2-5}
\end{equation}
which leads to
\begin{equation}
T_{c}=\frac{\eta}{9\sqrt{2} \pi q^{1/3}},
r_{c}=\frac{3 \sqrt{2} q^{1/3}}{\sqrt{\eta}},
P_{c}=\frac{\eta}{1152 \pi q^{2/3}}.
\label{3-2-6}
\end{equation}
From Eqs.(\ref{3-2-2}) and (\ref{3-2-3}), the pressure $P$ can be written as a function of $P(M,r_{+})$
\begin{equation}
P(M,r_{+})=\frac{2 \eta M -\eta r_{+}^{2}+6q^{2/3}\ln r_{+}}{16 \eta \pi r_{+}^{4}},
\label{3-2-7}
\end{equation}
the temperature $T$ can be written as a function of $T(M,r_{+})$
\begin{equation}
T(M,r_{+})=\frac{4\eta M-3q^{3/2}- \eta r_{+}^{2}+12 q^{3/2}\ln r_{+}}{4 \eta \pi r_{+}^{2}}.
\label{3-2-8}
\end{equation}
According to Eqs. (\ref{3-2-7}) and (\ref{3-2-8}), we shown the isenthalpic curves of this black hole in the $T-P$ plane.
\begin{figure*}
\vspace*{2cm}       
\includegraphics[width=0.4\textwidth]{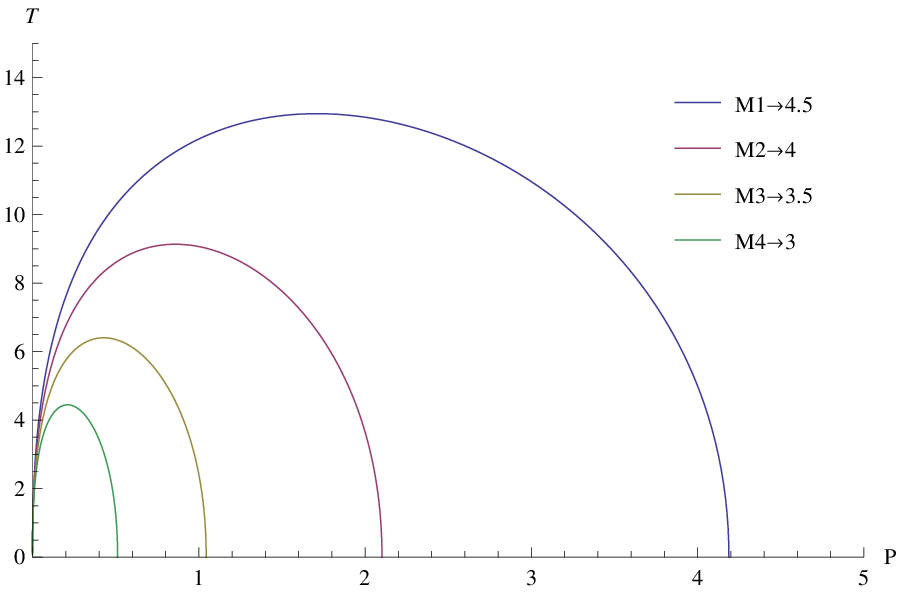}
\includegraphics[width=0.4\textwidth]{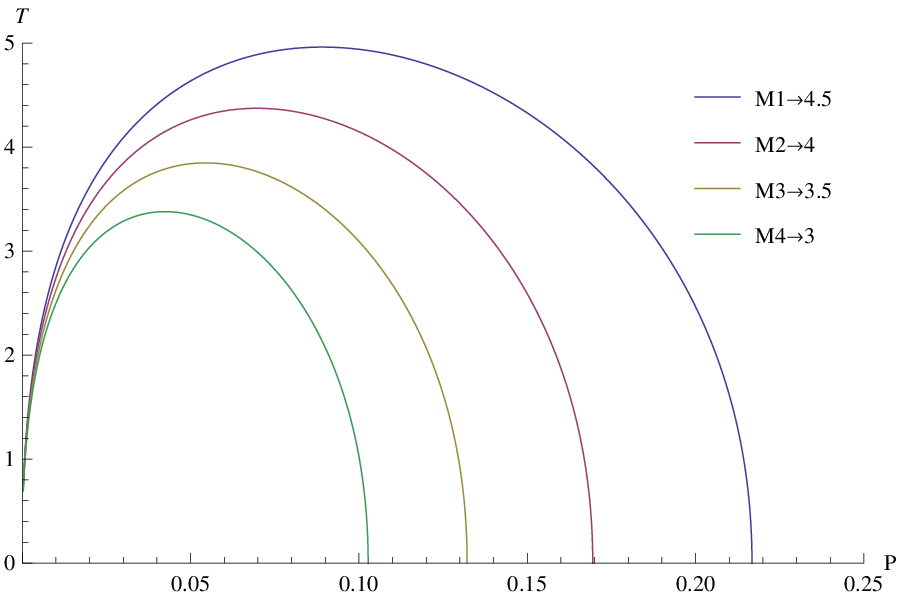}
\caption{The isenthalpic(constant mass) curves of a specific black hole in $f(R)$ gravity coupled with Yang-Mills field. From bottom to top, $M$=$3$, $3.5$, $4$, $4.5$, and the isenthalpic curves correspond to the increasing values of the mass $M$ of the black hole. The parameter $\eta=1$, the charge $q=1$(left) and $q=5$ (right).}
\label{fig:1}       
\end{figure*}
\par
In Fig.(\ref{fig:1}), the isenthalpic curves increases with the increase of the mass $M$ when the charge $q$ is a definite value, and it also decreases with the increase of the charge $q$ when the mass $M$ is a definite value. However, the temperature $T$ does not show monotonicity, which is increases/decreases with the change of the pressure $P$, and reverses at the highest point. The temperature corresponding to the inversion point can be called the inversion temperature. We consider the Joule-Thomson expansion to determine the hot and cold region and the inversion temperature.
\par
\begin{figure*}
\vspace*{4cm}       
\includegraphics[width=0.4\textwidth]{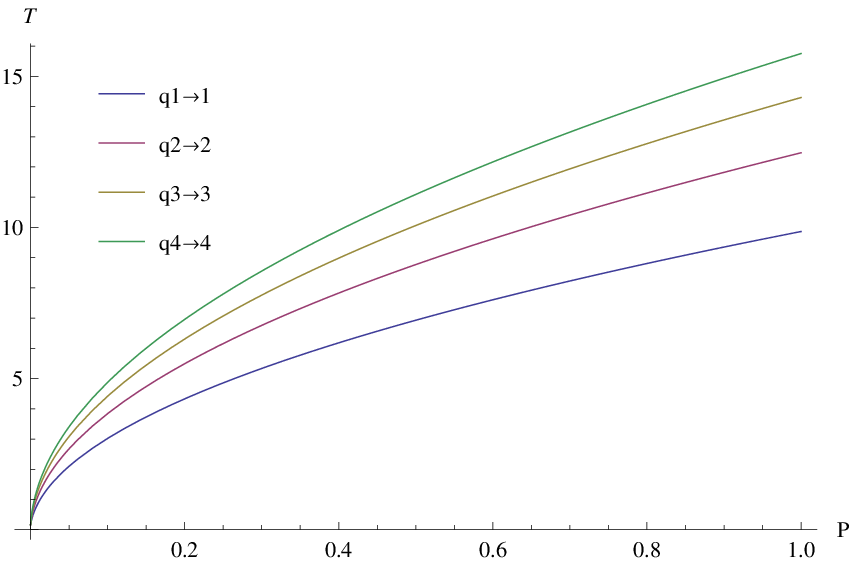}
\includegraphics[width=0.4\textwidth]{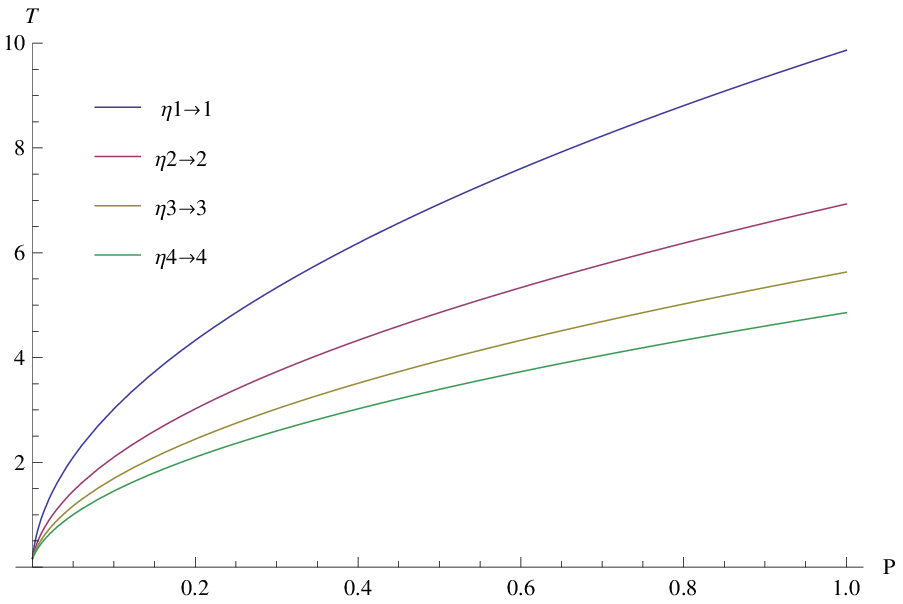}
\caption{The influence of the charge $q$ and the parameter $\eta$ on inversion temperature. The effect of the charge $q$ is shown on the left, the $\eta=1$, the $q=1, 2, 3, 4$ from bottom to top. The effect of the parameter $\eta$ is shown on the right, the charge $q=1$, the $\eta=1, 2, 3, 4$ from top to bottom.}
\label{fig:2}       
\end{figure*}

According to Eq.(\ref{3-1-1}), we can obtain
\begin{equation}
\mu=\Big(\frac{\partial T}{\partial P}\Big)_{M}=\Big(\frac{\partial T}{\partial r_{+}}\Big)_{M}\Big(\frac{\partial r_{+}}{\partial P}\Big)_{M}=\Big(\frac{\partial T}{\partial r_{+}}\Big)_{M}/\Big(\frac{\partial P}{\partial r_{+}}\Big)_{M}.
\label{3-2-9}
\end{equation}
Bring Eqs.(\ref{3-2-7}) and (\ref{3-2-8}) into the above expression, the Joule-Thomson coefficient is obtained, i.e.
\begin{equation}
\mu=\frac{2r_{+}\Big(-21q^{3/2}+\eta r_{+}^{2}(5+96P \pi r_{+}^{2})\Big)}{-3q^{3/2}+\eta r_{+}^{2}+32 \eta P \pi r_{+}^{4}},
\label{3-2-10}
\end{equation}
Setting $\mu=0$, we can get
\begin{equation}
-21q^{3/2}+\eta r_{+}^{2}(5+96P_{i} \pi r_{+}^{2})=0,
\label{3-2-11}
\end{equation}
and solve this equation for $r_{+}$ gives us four roots but only one root is physically meaningful, other roots are complex or negative. A positive and real root is
\begin{equation}
r_{+}=\frac{\sqrt{-\frac{5}{P_{i}\pi}+\frac{\sqrt{25\eta+8064 P_{i} \pi q^{3/2}}}{\sqrt{\eta}P_{i}\pi}}}{8\sqrt{3}},
\label{3-2-12}
\end{equation}
where the pressure $P_{i}$ is the inversion pressure. Then substituting the root into the temperature expression (\ref{3-2-3}), one can obtain the inversion temperature $T_{i}$. At the same times, we can also plot the inversion temperature curves in the $T-P$ plane, as shown in Fig(\ref{fig:2}).

\par
In Fig.(\ref{fig:2}), the inversion curves of a specific black hole in $f(R)$ gravity coupled with Yang-Mills field are presented for various values of the charge $q$ and the parameter $\eta$. We find that the inversion temperature of a given pressure increases with the increase of $q$. However, the parameter $\eta$ influence is different, which the inversion temperature decreases with the increase of $\eta$. On the other hand, we compare the inversion curve of this black hole with the inversion curve of the van der waals fluid, which is find that only a lower area and is an open curve, it means that the cooling region of this black hole is above the inversion curve on the Joule-Thomson expansion process.

\par
When $P_{i}=0$ in Eq.(\ref{3-2-11}), we can get the minimum inversion temperature $T_{i}^{min}$, i.e.
\begin{equation}
T_{i}^{min}=\frac{\sqrt{\frac{5 \eta}{21}}}{14 \pi q^{1/3}}.
\label{3-2-13}
\end{equation}
The ratio between the minimum inversion temperature and the critical temperature is given by
\begin{equation}
\frac{T_{i}}{T_{c}}\approx 0.4436.
\label{3-2-14}
\end{equation}
\begin{table}
\caption{Ratio between minimum inversion and critical temperature}
\label{tab:1}
\begin{center}
\begin{tabular}[t]{|l|c|c|c|}
  \hline
  Type  &  Ratio  &  Literature \\
  \hline
  $van-der-Waals-fluid$  & $0.75$  &  \cite{Ref39} \\
  $RN-AdS$  & $0.5$   &  \cite{Ref39} \\
  $Kerr-AdS$  &$0.5$  &   \cite{Ref40} \\
  $f(r)-gravity$  &  $0.5$   &  \cite{Ref44}\\
  $Bardeen-AdS$  &  $0.5366$ & \cite{Ref49}\\
  \hline

\end{tabular}
\end{center}
\end{table}
%
%
\par
\begin{figure*}
\vspace*{4cm}       
\includegraphics[width=0.4\textwidth]{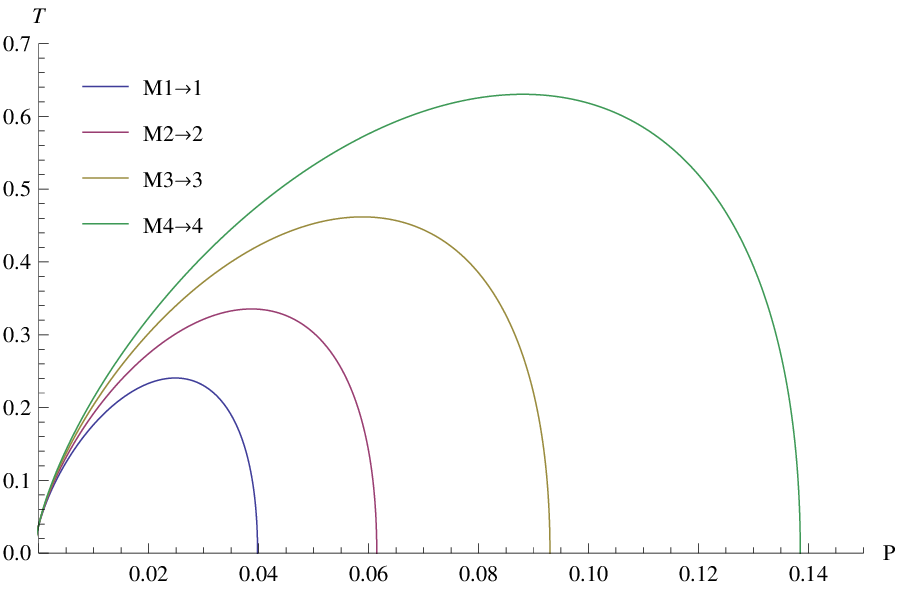}
\includegraphics[width=0.4\textwidth]{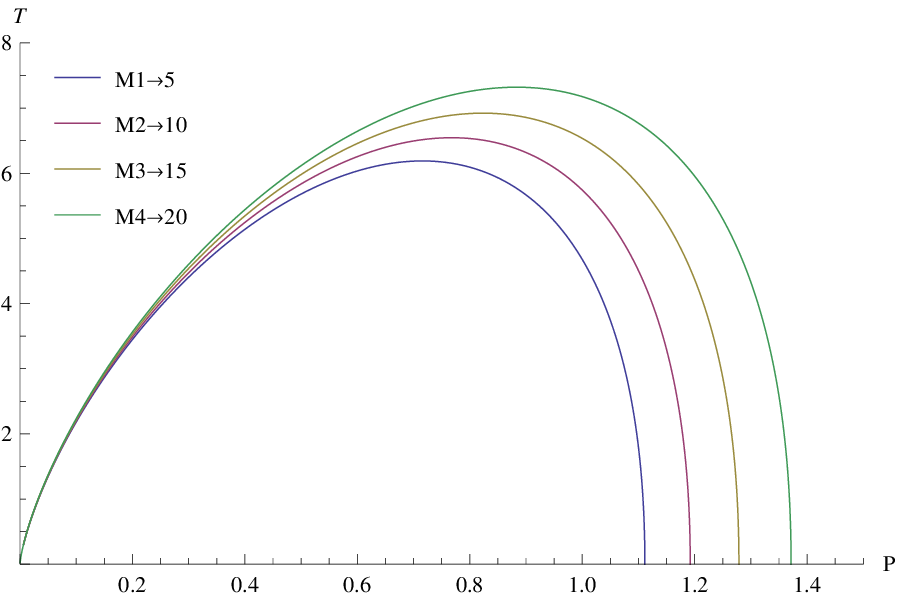}
\caption{The isenthalpic(constant mass) curves of a specific black hole in $f(R)$ gravity coupled with Yang-Mills field. From bottom to top, $M$=$1$, $2$, $3$, $4$(left), $M$=$5$, $10$, $15$, $20$ and the isenthalpic curves correspond to the increasing values of the mass $M$ of the black hole. The parameter $\eta=1$, the charge $q=1$(left) and $q=5$ (right).}
\label{fig:3}       
\end{figure*}
\begin{figure*}
\vspace*{4cm}       
\includegraphics[width=0.4\textwidth]{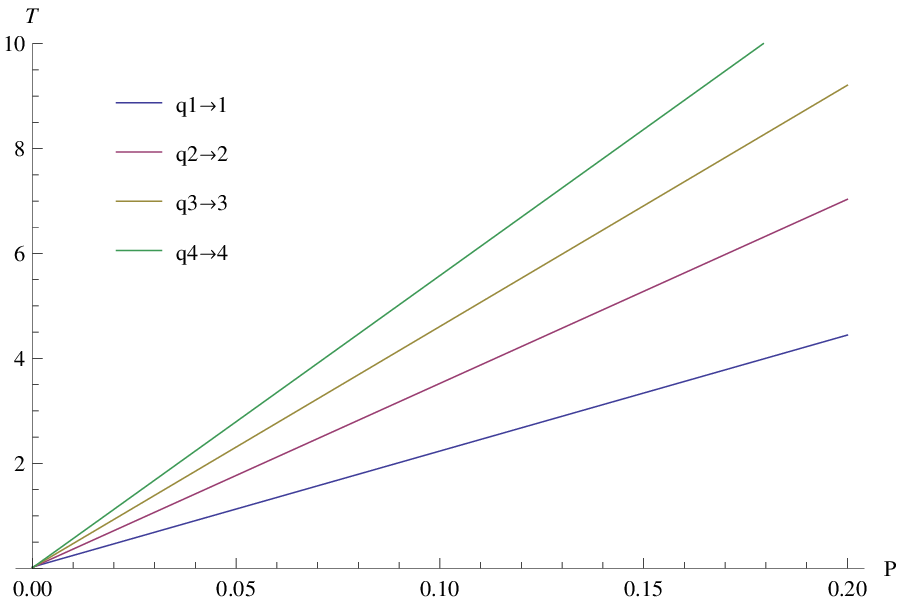}
\includegraphics[width=0.4\textwidth]{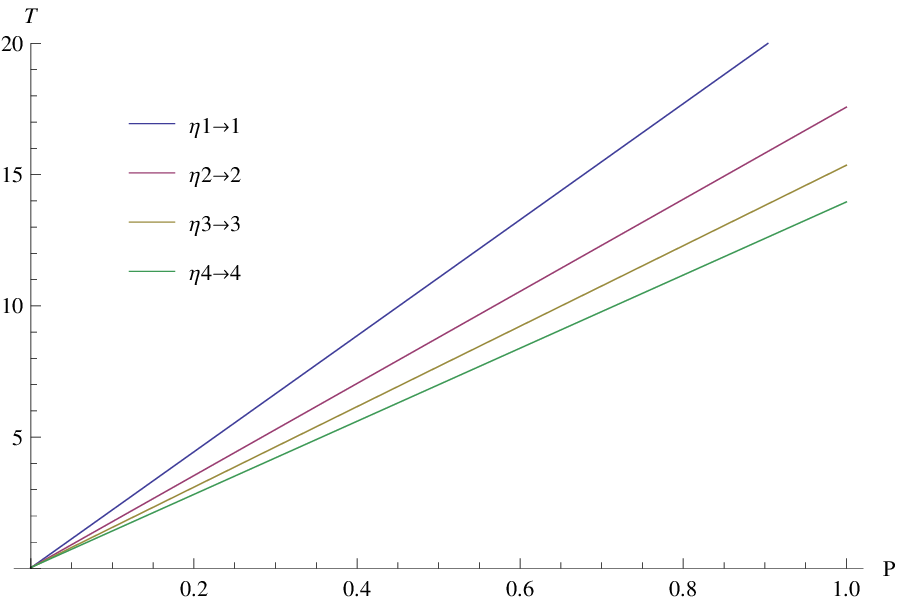}
\caption{The influence of the charge $q$ and the parameter $\eta$ on inversion temperature. The effect of the charge $q$ is shown on the left, the $\eta=1$, the $q=1, 2, 3, 4$ from bottom to top. The effect of the parameter $\eta$ is shown on the right, the charge $q=1$, the $\eta=1, 2, 3, 4$ from top to bottom.}
\label{fig:4}       
\end{figure*}

When $d=5$, the metric function $f(r)$ is given by Eq.(\ref{2-5})
\begin{equation}
f(r)=\frac{2}{3}+8\pi P r^{2}-\frac{M}{r^{3}}-\frac{72 q^{2}\ln{r}}{5\eta r^{3}},
\label{3-2-15}
\end{equation}
the equation of state is
\begin{equation}
P=\frac{T}{10r_{+}}-\frac{1}{20\pi r_{+}^{2}}+\frac{9q^{3}}{25\eta \pi r_{+}^{5}}.
\label{3-2-16}
\end{equation}
According Eq.(\ref{3-2-5}), one can get
\begin{equation}
T_{c}=\frac{\sqrt[3]{3\eta}}{8\pi q^{2/3}},
r_{c}=\frac{2{(3q)}^{2/3}}{\sqrt[3]{\eta}},
P_{c}=\frac{{\eta}^{2/3}}{400\pi \sqrt[3]{3q}}.
\label{3-2-17}
\end{equation}
We also obtain the pressure $P$ as a function of $P(M,r_{+})$
\begin{equation}
P(M,r_{+})=\frac{15 \eta M-10\eta r_{+}^{3}+216 q^{2} \ln{r_{+}}}{120 \eta \pi r_{+}^{5}},
\label{3-2-18}
\end{equation}
and the temperature $T$ as a function of $T(M,r_{+})$
\begin{equation}
T(M,r_{+})=\frac{75\eta M-216 q^{2}-20\eta r_{+}^{3}+1080q^{2}\ln{r_{+}}}{60\eta \pi r_{+}^{4}}.
\label{3-2-19}
\end{equation}
The Joule-Thomson coefficient is
\begin{equation}
\mu=\frac{-648q^{2}r_{+}+20\eta r_{+}^{4}(3+40 P \pi r_{+}^{2})}{-36 q^{2}+5\eta r_{+}^{3}(1+20 P \pi r_{+}^{2})},
\label{3-2-20}
\end{equation}
and the minimum inversion temperature $T_{i}^{min}$ is
\begin{equation}
T_{i}^{min}=\frac{\sqrt[3]{5\eta}}{18\sqrt[3]{2}\pi q^{2/3}}.
\label{3-2-21}
\end{equation}
The ratio between the minimum inversion temperature and the critical temperature is given by
\begin{equation}
\frac{T_{i}}{T_{c}}\approx 0.418238.
\label{3-2-22}
\end{equation}
We also plot the isenthalpic curves and the inversion curves of this black hole in the five-dimensional case, as shown in Figs.(\ref{fig:3}) and (\ref{fig:4}). At the same time, we find that the isenthalpic curves and the inversion curves are similar to the four-dimensional cases.
\par
However, we find that $\frac{T_{i}}{T_{c}}$ is different. Based on this consider, we investigate this ratio in higher dimensions. Therefore, we plot the inversion curves and the isenthalpic curves of this black hole in six and seven dimensions, and the ratio between the minimum inversion temperature and the critical temperature is obtained. The equation of state is
\begin{eqnarray}
&&P_{6}=\frac{T}{12r_{+}}-\frac{1}{16\pi r_{+}^{2}}+\frac{5\sqrt{2}\sqrt[4]{3}q^{5/2}}{6\eta \pi r_{+}^{6}}(d=6),\nonumber\\
&&P_{7}=\frac{7T}{98r_{+}}-\frac{7}{98\pi r_{+}^{2}}+\frac{375q^{2}}{49\eta \pi r_{+}^{7}}(d=7).
\label{3-2-23}
\end{eqnarray}
According Eq. (\ref{3-2-5}), we can get
\begin{eqnarray}
&&T_{6c}=\frac{\sqrt[8]{2}3^{15/16}\sqrt[4]{\eta}}{5\sqrt{5}\pi q^{5/8}}(d=6),\nonumber\\
&&T_{7c}=\frac{5^{2/5}\sqrt[5]{\eta}}{3^{5/7}\sqrt[5]{2}\pi q^{2/5}}(d=7).
\label{3-2-24}
\end{eqnarray}
We can change the pressure $P$ and temperature $T$ into $P(M,r_{+})$ and $T(M,r_{+})$ forms, for $d=6$
\begin{eqnarray}
&&P_{6}(M,r_{+})=\frac{4\eta M-3 \eta r_{+}^{4}+160\sqrt{2}\sqrt[4]{3}q^{5/2}\ln{r_{+}}}{32\eta \pi r_{+}^{6}},\nonumber\\
&&T_{6}(M,r_{+})=\frac{12\eta M-80\sqrt{2}\sqrt[4]{3}q^{5/2}-3\eta r_{+}^{4}+480\sqrt{2}\sqrt[4]{3}q^{5/2}\ln{r_{+}}}{8\eta \pi r_{+}^{5}},
\label{3-2-25}
\end{eqnarray}
for $d=7$
\begin{eqnarray}
&&P_{7}(M,r_{+})=\frac{35\eta M-28\eta r_{+}^{5}+15000q^{3}\ln{r_{+}}}{280\eta \pi r_{+}^{7}},\nonumber\\
&&T_{7}(M,r_{+})=\frac{245\eta M-15000q^{3}-56\eta r_{+}^{5}+105000q^{3}\ln{r_{+}}}{140\eta \pi r_{+}^{6}}.
\label{3-2-26}
\end{eqnarray}
The Joule-Thomson coefficient is
\begin{eqnarray}
&&\mu_{6}=\frac{-880\sqrt{2}\sqrt[4]{3}q^{5/2}r_{+}+6\eta r_{+}^{5}(7+80P \pi r_{+}^{2})}{-40\sqrt{2}\sqrt[4]{3}q^{5/2}+3\eta r_{+}^{4}+48\eta P \pi r_{+}^{6}}(d=6),\nonumber\\
&&\mu_{7}=\frac{4r_{+}\Big(-4875q^{3}+14\eta r_{+}^{5}(2+21P \pi r_{+}^{2})\Big)}{-750q^{3}+7\eta r_{+}^{5}+98\eta P \pi r_{+}^{7}}(d=7).
\label{3-2-27}
\end{eqnarray}
The minimum inversion temperature is
\begin{eqnarray}
&&T_{6i}^{min}=\frac{3^{19/16}\sqrt[4]{7\eta}}{11\sqrt[4]{55}2^{7/8}\pi q^{5/8}}(d=6),\nonumber\\
&&T_{7i}^{min}=\frac{\sqrt[5]{\frac{7}{39}}10^{2/5}\sqrt[5]{\eta}}{13\pi q^{2/5}}(d=7).
\label{3-2-28}
\end{eqnarray}
\begin{figure*}
\vspace*{4cm}       
\includegraphics[width=0.4\textwidth]{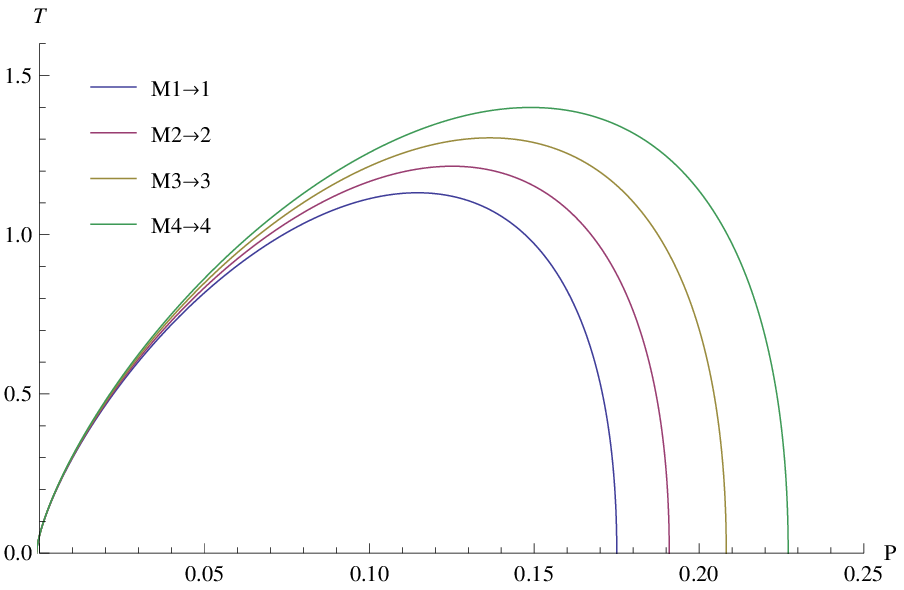}
\includegraphics[width=0.4\textwidth]{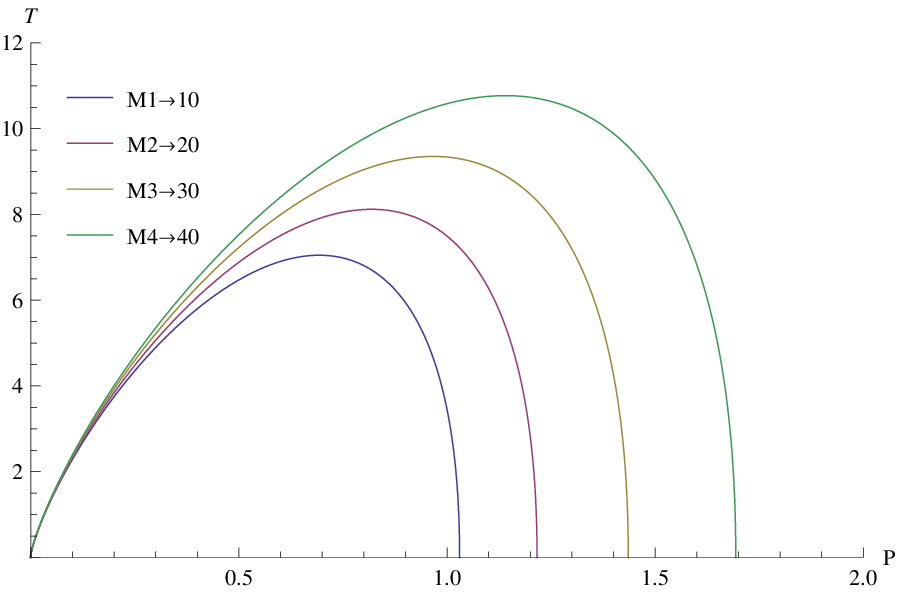}
\caption{Constant mass isenthalpic curves, two graphs corresponding to dimensions $d=6$ (left) and $d=7$ (right). When $d=6$, the mass $M$=$1$, $2$, $3$, $4$ from bottom to top. When $d=7$, the mass $M$=$10$, $20$, $30$, $40$ from bottom to top, the parameter $\eta=1$, the charge $Q=1$.}
\label{fig:5}       
\end{figure*}
\begin{figure*}
\vspace*{4cm}       
\includegraphics[width=0.4\textwidth]{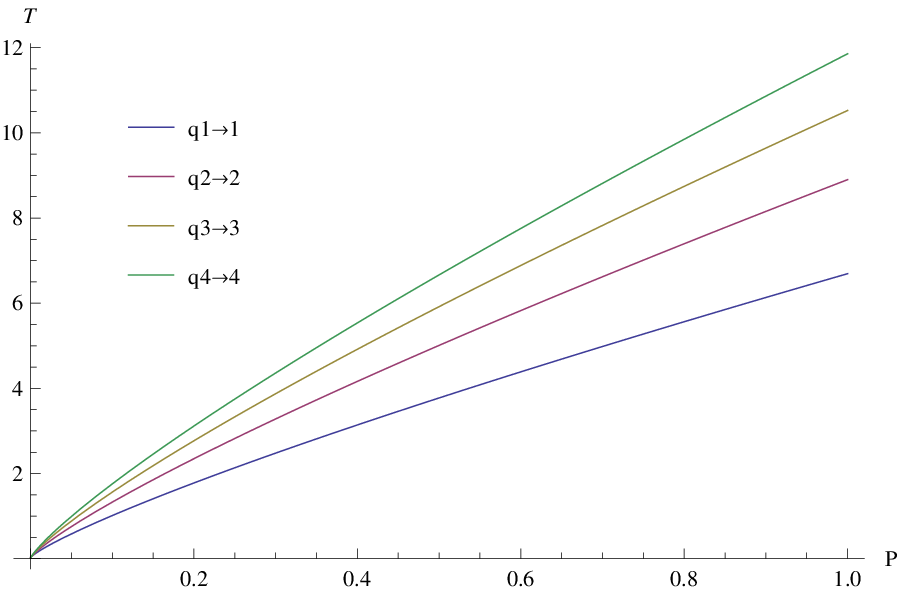}
\includegraphics[width=0.4\textwidth]{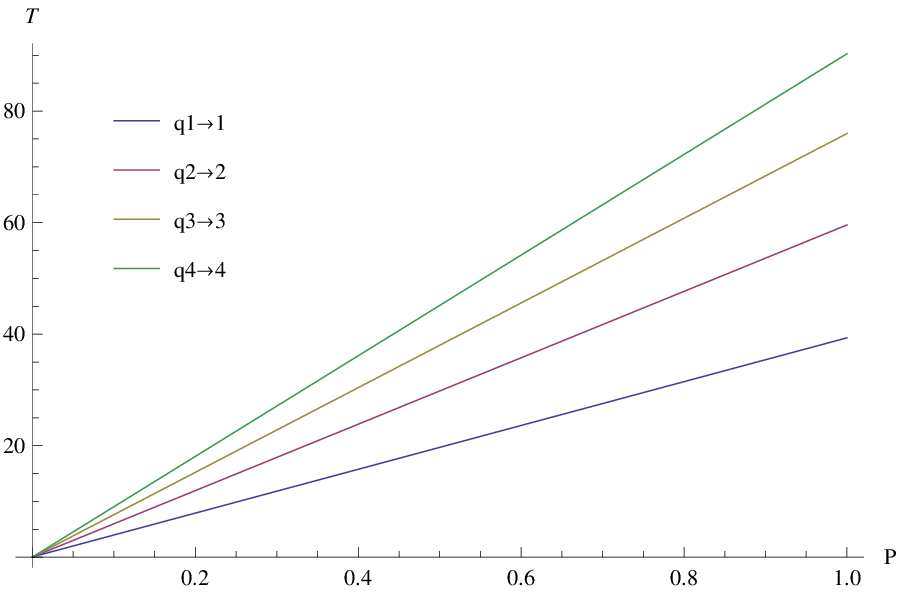}
\caption{Inversion curves, (5a) is the inversion curves in $d=6$, (5b) is the inversion curves in $d=7$. The charge $q$=$1$, $2$, $3$, $4$, the parameter $\eta=1$.}
\label{fig:6}       
\end{figure*}
We plot the isenthalpic curves and the inversion curves in the $T-P$ plane, as shown in Figs.(\ref{fig:5}) and (\ref{fig:6}). The radio of the minimum inversion temperature and the critical temperature is
\begin{eqnarray}
&&\frac{T_{6i}^{min}}{T_{6c}}\approx 0.399481(d=6),\nonumber\\
&&\frac{T_{7i}^{min}}{T_{7c}}\approx 0.384911(d=7).
\label{3-2-29}
\end{eqnarray}
\begin{table}
\caption{\bf Ratio between minimum inversion and critical temperature in different dimension}
\label{Tab:2}
\begin{center}
\setlength{\tabcolsep}{2mm}
\linespread{3}
\begin{tabular}[t]{|l|c|c|c|c|}
  \hline
  Dimension  &  $d=4$   &  $d=5$  &  $d=6$   &  $d=7$ \\
   \hline
   Ratio  &  $0.4436$  & $0.418238$  & $0.399481$  &$0.384911$ \\
  \hline
\end{tabular}
\end{center}
\end{table}

In this section, we discuss the Joule-Thomson expansion in high dimensions and get the ratio of between the minimum inversion and the critical temperature, the inversion curves, the isenthalpic curves, and found that the physical characteristics are similar to those of low-dimensional cases.
%
%
\section{Conclusions}
\label{sec4}
%
%
%
\par\noindent
In this paper, we have investiagted Joule-Thomson expansion of a specific black hole in $f(R)$ gravity coupled with Yang-Mills field in different dimensions. The Joule-Thomson expansion is a phenomenon of temperature change when gas expands irreversibly from high pressure to low pressure through porous plugs or valves. $f(R)$ gravity is an important correction of Einstein gravity, and the Yang-Mills field is one of the interesting non-abelian gauge theory.

\par
Firstly, we reviewed the thermodynamic properties of a specific black hole in $f(R)$ gravity coupled with Yang-Mills field, and given the equation of state, the first law of thermodynamics and the Smarr relation. Then, we investigated the Joule-Thomson expansion of this black hole in different dimensions, and mainly discussed the inversion curves, the isenthalpic curves and the inversion temperature. Because the enthalpy remains constant in the Joule-Thomson expansion process. The mass can be interpreted as enthalpy in the extended phase space. Therefore, we can describe the isenthalpic curves by determining the mass of the black hole. When $d=4$, the isenthalpic curves of a given pressure increases with the increase of $M$, and the isenthalpic curves of a given mass decrease with the increase of $q$. In particular, the temperature change reverses at the highest point. We got the inversion curves by Joule-Thomson expansion, and according to the positive and negative of the Joule-Thomson coefficient $\mu$, we can known that above the inversion curve is the cooling region of the black hole and on the other side is the heating region. The inversion curve receives the joint influence of the charge $q$ and the parameter $\eta$, which the inversion curves increases gradually with the increase of $q$, but the effect of parameter $\eta$ is quite different from that of $q$, the inversion curves decreases gradually with the increase of $\eta$.

\par
On the other hand, we found that the radio of the minimum inversion temperature and the critical temperature of this black hole is smaller than the van der Waals system, and compared with RN-AdS black hole, regular(Bardeen)-AdS black hole and the charge of black hole in $f(R)$ gravity, this ratio is smaller, so we consider this is due to the couple of the Yang-Mills field. Finally, we investigated the inversion curves and the isenthalpic curves of $d$= $5$, $6$, $7$ dimensions in order to obtaind the effect of dimension on this process, and also obtained the radio of the minimum inversion temperature and the critical temperature. We found that the ratio decreases gradually with the increase of dimensions, and the parameter $\eta$ has no effect on this ratio.
%
%
\section{Acknowledgements}
\par\noindent
The authors would like to thank the anonymous reviewers for their helpful comments and suggestions, which helped to improve the quality of this paper. This work was supported by the National Natural Science Foundation of China No. 11903025, by the Program for NCET-12-1060, by the Sichuan Youth Science and Technology Foundation with Grant No. 2011JQ0019, and by FANEDD with Grant No. 201319, and by the Innovative Research Team in College of Sichuan Province with Grant No. 13TD0003, and by Ten Thousand Talent Program of Sichuan Province, and by Sichuan Natural Science Foundation with Grant No. 16ZB0178, and by the starting funds of China West Normal University with Grant No.17YC513 and No.17C050.
%
%

%
%

\begin{thebibliography}{99}
%
%
\bibitem{Ref1}
S. W. Hawking, Nature \textbf{248}, (1974) 30.

\bibitem{Ref2}
J. D. Bekenstein, Phys. Rev. D \textbf{7}, (1973) 2333.

\bibitem{Ref3}
J. M. Bardeen, B. Carter, and S. W. Hawking, Commun. Math. Phys \textbf{31} (1973) 161.

\bibitem{Ref4}
S. W. Hawking, Commun. Math. Phys \textbf{43}, (1975) 199.

\bibitem{Ref5}
S. W. Hawking, D. N. Page, Commun. Math. Phys \textbf{87}, (1983) 577.

\bibitem{Ref6}
A. Chamblin, R. Emparan, C. V. Johnson, R. C. Myers, Phys. Rev. D \textbf{60}, (1999) 064018.

\bibitem{Ref7}
A. Chamblin, R. Emparan, C. V. Johnson, R. C. Myers, Phys. Rev. D. \textbf{60}, (1999) 104026.

\bibitem{Ref8}
Cvetic. M, Gubser. S. S, JHEP \textbf{437}, (1999) 4.

\bibitem{Ref9}
Niu. C, Tian. Y, Wu. X, Phys. Rev. D. \textbf{84} (2011), 025017.

\bibitem{Ref10}
B. P. Dolan, Class. Quant. Grav. \textbf{28} (2011), 235017.

\bibitem{Ref11}
D. Kastor, S. Ray, and J. Traschen, Class. Quant. Grav. \textbf{26} (2009) 195011.

\bibitem{Ref12}
B. Dolan, Class. Quant. Grav \textbf{28}, (2011) 125020.

\bibitem{Ref13}
B. P. Dolan, Phys. Rev. D. \textbf{84}, (2011) 127503.

\bibitem{Ref14}
M. Cvetic, G. Gibbons, D. Kubiznak, and C. Pope, Phys. Rev. D. \textbf{84}, (2011) 024037.

\bibitem{Ref15}
H. Lu, Y. Pang, C. N. Pope, and J. F. Vazquez-Poritz, Phys. Rev. D. \textbf{86}, (2012) 044011.

\bibitem{Ref16}
Chen. S, Liu. X, Liu. C, Chinese. Physics. Letters. \textbf{6}, (2013) 060401.

\bibitem{Ref17}
R. G. Cai, Li, et al, JHEP \textbf{22}, (2013) 9.

\bibitem{Ref18}
Belhaj, A, et al, EPJC \textbf{76}, (2016) 73.

\bibitem{Ref19}
Bhattacharya. K, Majhi. B. R, Samanta. S, Phys. Rev. D \textbf{96}, (2017) 084037.

\bibitem{Ref20}
Z. X. Xiong, L. F. Ling, Advances in High Energy Physics \textbf{12}, (2016) 2016.

\bibitem{Ref21}
J. X. Mo, G. Q. Li, X. B. Xu, EPJC \textbf{26}, (2016) 545.

\bibitem{Ref22}
Y. G. Miao, Y. M. Wu, Advances in High Energy Physics. \textbf{12}, (2017).

\bibitem{Ref23}
Mandal. A, Samanta. S, Majhi. B. R, Phys. Rev. D. \textbf{94}, (2016) 064069.

\bibitem{Ref24}
Dehyadegari. A, Sheykhi. A, Montakhab. A, Phys. Letters. B. \textbf{235}, (2016) 768.

\bibitem{Ref25}
Hendi. S. H, Panahiyan. S, Panah. B. E, et al, Phys. Rev. D. \textbf{92}, (2016) 024028.

\bibitem{Ref26}
Upadhyay. S, Pourhassan. B, Farahan. i. H, Phys. Rev. D. \textbf{95}, (2017) 106014.

\bibitem{Ref27}
Sadeghi. J, Pourhassan. B, Rostami. M, Phys. Rev. D. \textbf{94}, (2016) 064006.

\bibitem{Ref28}
Cheng. P, Wei. S. W, Liu. Y. X, Phys. Rev. D. \textbf{94}, (2016) 2.

\bibitem{Ref29}
Guo. X, Li. H, Zhang. L, et al, Advances in High Energy Physics \textbf{1}, (2016) 10.

\bibitem{Ref30}
J. X. Mo, G. Q. Li and X. B. Xu, Phys. Rev. D. \textbf{93}, (2016) 084041.

\bibitem{Ref31}
Wei. S. W, Cheng. P, Liu. Y. X, Phys. Rev. D. \textbf{93}, (2015) 084015.

\bibitem{Ref32}
J. Xu, L. M. Cao and Y. P. Hu, Phys. Rev. D. \textbf{91}, (2015) 124033.

\bibitem{Ref33}
M. H. Dehghani, S. Kamrani and A. Sheykhi, Phys. Rev. D. \textbf{90}, (2014) 104020.

\bibitem{Ref34}
Zou. D. C, Zhang. S. J, Wang. B, Phys. Rev. D. \textbf{89}, (2014) 4.

\bibitem{Ref35}
Kuang. X. M, Miskovic. O, Phys Rev. D. \textbf{95}, (2017) 046009.

\bibitem{Ref36}
Kuang. X. M, Wu. J. P, EPJC \textbf{77}, (2017) 670.

\bibitem{Ref37}
S. H. Hendi, S. Panahiyan, B. Eslam. Panah, Int. J. Mod. Phys. D. \textbf{25}, (2016) 1650010.

\bibitem{Ref38}
S. H. Hendi, R. B. Mann, S. Panahiyan, B. Eslam Panah, Phys. Rev. D. \textbf{95}, (2017) 021501.

\bibitem{Ref39}
\"{O}zg\"{u}r. \"{O}kc\"{u}, Ekrem. Ayd{\i}ner, EPJC \textbf{77}, (2017) 24 .

\bibitem{Ref40}
\"{O}zg\"{u}r \"{O}kc\"{u}, Ekrem Ayd{\i}ner, EPJC \textbf{78} (2018) 123.

\bibitem{Ref41}
J. X. Mo, G. Q. Li, S. Q, Lan, X. B. Xu, Phys. Rev. D. \textbf{98}, (2018) 124032 .

\bibitem{Ref42}
H. Ghaffarnejad, E. Yaraie, M. Farsam, Int. J. Theo. Phys \textbf{57}, (2018) 1671.

\bibitem{Ref43}
R. D. Almeida, K. P. Yogendran, arXiv:1802.05116.

\bibitem{Ref44}
M. Chabab, H. El Moumni, S. Iraoui, K. Masmar, S. Zhizeh, LHEP \textbf{2}, (2018) 02.

\bibitem{Ref45}
C. L. Ahmed Rizwan, A. Naveena Kumara, Deepak. Vaid, K. M. Ajith, Int. J. Mod. Phys. \textbf{33}, (2018) 1850210.

\bibitem{Ref46}
J. X. Mo, G. Q. Li, arXiv:1805.04327.

\bibitem{Ref47}
S. Q. Lan, Phys. Rev. D. \textbf{98}, (2014) 084014.

\bibitem{Ref48}
A. Cisterna, S. Q. Hu, X. M. Kuang, arXiv:1808.07392.

\bibitem{Ref49}
J. Pu, S. Guo, Q. Q. Jiang, X. T. Zu, arXiv:1905.02318.

\bibitem{Ref50}
D. M. Yekta, A. Hadikhani, \"{O}. \"{O}kc¨¹, arXiv:1905.03057.

\bibitem{Ref51}
S. Guo, J. Pu, Q. Q. Jiang, arXiv:1905.03604.

\bibitem{Ref52}
Starobinsky. A. A, Physics. Letters. B  \textbf{91}, (1980) 1.

\bibitem{Ref53}
Felice. A. D, Tsujikawa. S, Living Reviews in Relativity. \textbf{13}, (2010) 3.

\bibitem{Ref54}
Nojiri. S, Odintsov. S. D, Phys. Rept. \textbf{505} (2011) 2.

\bibitem{Ref55}
S. Capozziello, M. D. Laurentis, arXiv:1108.6266.

\bibitem{Ref56}
P. B. Yasskin, Phys. Rev. D  \textbf{12} (1975) 2212.

\bibitem{Ref57}
M. S. Volkov and D. V. Galtsov, Phys. Rept. \textbf{319} (1999) 83.

\bibitem{Ref58}
S. Habib. Mazharimousavi, M. Halilsoy, Phys. Rev. D. \textbf{84} (2011) 064032.

\bibitem{Ref59}
D. Kubiznak, R. B. Mann, J. High. Energy. Phys. \textbf{07} (2012) 033.
%
\end{thebibliography}
\end{document}